\documentclass[%
 reprint,
 amsmath,amssymb,
 aps,
]{revtex4-2}
\usepackage{graphicx}
\usepackage{dcolumn}
\usepackage{bm}
\usepackage[usenames]{color}
\usepackage{colortbl}
\usepackage{hyperref}
\begin{document}


\title{Accurate determination of the Josephson critical current by lock-in measurements}

\author{Razmik A. Hovhannisyan$^{1,2}$}
\email{Razmik.Hovhannisyan@fysik.su.se}
\author{Olena M. Kapran$^1$}
\author{Taras Golod$^1$}
\author{Vladimir M. Krasnov$^{1,2}$}
\email{Vladimir.Krasnov@fysik.su.se} \affiliation{$^1$ Department
of Physics, Stockholm University, AlbaNova University Center,
SE-10691 Stockholm, Sweden} \affiliation{$^2$ Moscow Institute of
Physics and Technology (State University), 9 Institutskiy per.,
141700 Dolgoprudny, Russia}

\date{\today}
\begin{abstract}
Operation of Josephson electronics usually requires determination of the Josephson critical current $I_c$, which is affected both by fluctuations and measurement noise. Lock-in measurements allow obviation of $1/f$ noise and, therefore, provide a major advantage in terms of noise and accuracy with respect to conventional dc-measurements. In this work we show both theoretically and experimentally that the $I_c$ can be accurately extracted using first and third harmonic lock-in measurements of junction resistance. We derive analytic expressions and verify them experimentally on nano-scale Nb-PtNi-Nb and Nb-CuNi-Nb Josephson junctions. 
\end{abstract}

\maketitle

\newpage

\section{INTRODUCTION}

A Josephson junction (JJ) is the key element of superconducting electronics \cite{Likharev}. Operation of Josephson devices usually involves manipulation and determination of the Josephson critical current, $I_c$. Conventional dc-measurements of $I_c$ are complicated by two factors. First, $I_c$ in small junctions is subject to both thermal and quantum fluctuations \cite{Martinis_1987,Grabert_1988,Kautz_1990,Krasnov_2007}. The latter are particularly large in quantum devices, such as qubits, and require statistical determination of $I_c$ with a large number of measurements \cite{Martinis_1987,Krasnov_2007,Makhlin_2000,Fazio_2002}. Fluctuations are significant even for classical devices containing small JJ's, such as sensors \cite{Pekola_2014}, nano-SQUIDs \cite{Gross_1990,Koelle_1999,Kirtley_2016,Zeldov_2010,Gupta_2018}, low-dissipation digital electronics \cite{Likharev,Dimov_2008}, as well as for JJ's used in fundamental studies of unconventional superconductors \cite{Wiedenmann_2016,Dominguez_2012,Kalenyuk_2018}.
Second, dc-measurements are strongly affected by the flicker $1/f$ noise. Fluctuations and noise together could lead to smearing of the current-voltage ($I$-$V$) characteristics of JJ's \cite{Kautz_1990} and make $I_c$ an ill-defined quantity. Lock-in measurements at high enough frequency facilitate obviation of the $1/f$ noise. Simultaneously, they allow a statistical averaging over an arbitrary number of periods. In recent works \cite{Iovan_2014,Kapran_2020} it has been noticed that the magnetic field modulation of the junction lock-in resistance reflects the corresponding $I_c(H)$ modulation and can be used for extraction of $I_c$. However, such extraction requires proper mathematical justification and experimental verification, which is the main motivation of this work.   

In this work we study both theoretically and experimentally how the critical current of resistively shunted Josephson junctions (RSJ) can be deduced from lock-in measurements of ac-resistance, $R_{ac}$. First we present a simple analytical solution for the relation between $I_c$ and different harmonics of $R_{ac}$. Next, we use derived expressions for determination of $I_c$ for nano-scale proximity-coupled Nb-PtNi-Nb and Nb-CuNi-Nb JJ's. We demonstrate that the formalism leads to a robust reconstruction of $I_c$ in a broad range of ac-current amplitudes, $I_{ac}$. We also show that, with some minor adjustments taking into account eventual field-dependence of the normal resistance, $R_n(H)$, and deviations of the $I$-$V$ shape from the RSJ model, the formalism can be employed for accurate determination of the $I_c(H)$ modulation. We conclude that it is advantageous to use both the 1st and the 3rd lock-in harmonics for unambiguous determination of $I_c$.    
   
\section{Theoretical analysis of the lock-in response in the RSJ model}
The shape of the $I$-$V$ in the RSJ model is
\begin{equation}
    V = I R_n \sqrt{1 - (I_c/I)^2}
\end{equation}
for $I>I_c$ and $V=0$ for $I<I_c$. We assume that the bias is provided by the periodic ac-current, $I = I_{ac}\sin{\omega t}$, with the period $T=2\pi/\omega$ and the amplitude $I_{ac}>I_c$. The $m$-{ th} harmonics of the lock-in response at $\omega_m=m\omega$ is given by the $m$-{ th} Fourier component, 
\begin{equation}
    V_m = \frac{1}{T}\int_{-T/2}^{T/2} V(t) \sin(m \omega t) dt,
\end{equation}
Eqs. (1) { and} (2) lead to simple expressions for lock-in harmonics of resistance, $R_m = V_m/I_{ac}$, first three of which are:
\begin{eqnarray}
\frac{R_1}{R_n}=1-\left[\frac{I_c}{I_{ac}}\right]^2,\\
R_2=0,\\
\frac{R_3}{R_n}=\left[\frac{I_c}{I_{ac}}\right]^4-\left[\frac{I_c}{I_{ac}}\right]^2.
\end{eqnarray}
Thus the $I_c$ can be deduced from either 1st or 3rd harmonics of the lock-in resistance: 
\begin{eqnarray}
I_c(R_1) =  I_{ac}\sqrt{1-\frac{R_1}{R_n}},\\
I_c(R_3) = \frac{I_{ac}}{\sqrt{2}}\sqrt{1 - \sqrt{1+4\frac{R_3}{R_n}}}
\end{eqnarray}

\begin{figure*}[!ht]
    \begin{center}
        \includegraphics[width=14cm]{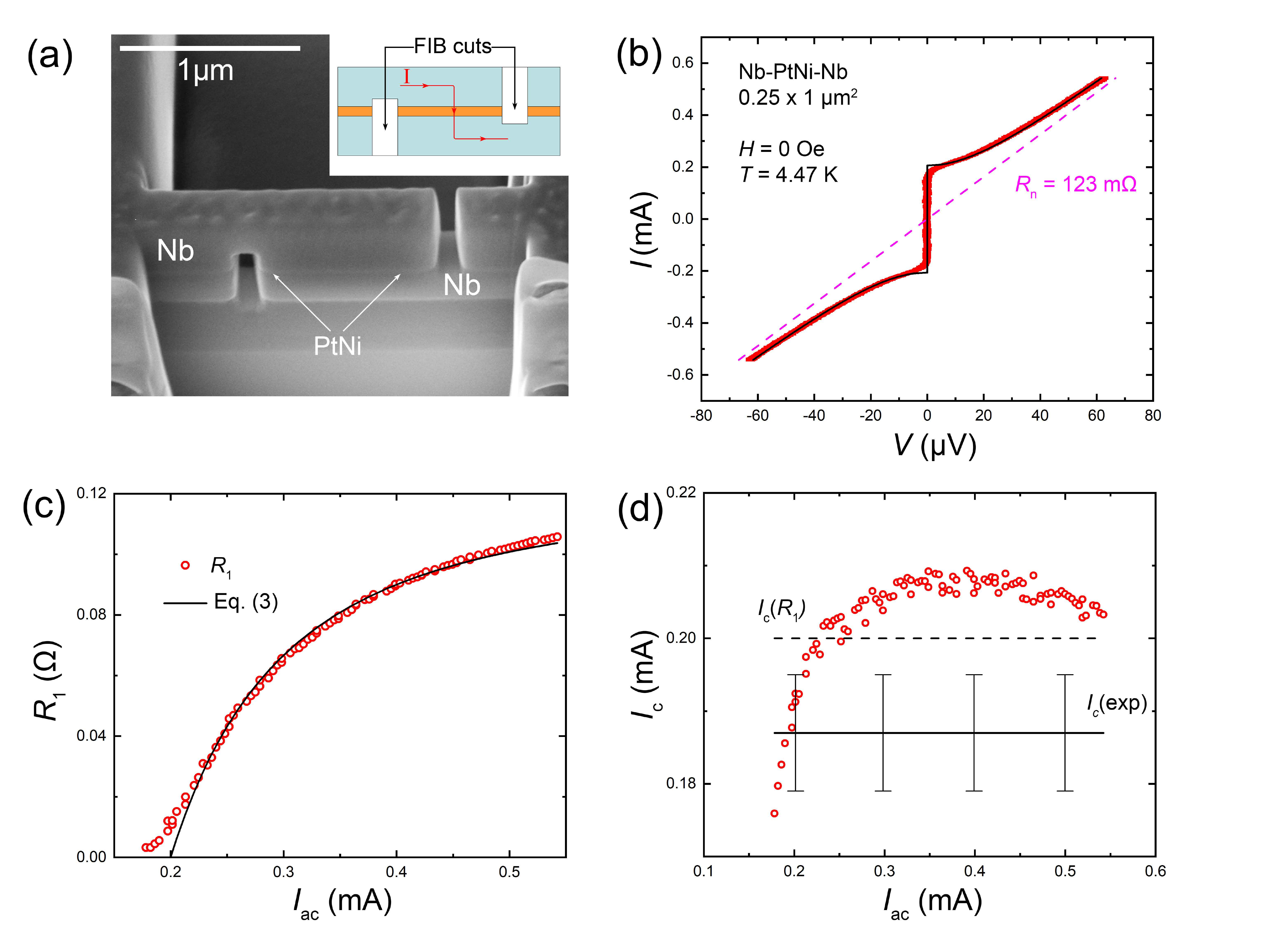}
        \caption{(a) SEM image and a sketch of a Nb-PtNi-Nb junction. (b) Experimental current-voltage characteristics of the junction at $H=0$ and $T=4.47$ K (red symbols). The black line represents the RSJ fit, Eq. (1). (c) Red circles show the dependence of the 1st harmonic lock-in resistance on the ac-current amplitude for $H=0$ and $T=4.47$ K. The black line shows the fit by Eq. (3), using $I_c$ as a fitting parameter. (d) Red circles represent $I_c$ vs. $I_{ac}$ reconstructed from the data in (c), using Eq. (6) with $R_n$ as a fitting parameter. The solid horizontal line represents $I_c({\textrm{exp}})$ obtained from the $I$-$V$ in (b) with error bars due to smearing at $I\simeq I_c$. The dashed line represents $I_c(R_1)$ obtained from the fitting by Eq. (3) in (c). A small systematic overestimation of the reconstructed $I_c$ is caused by smearing of the experimental $I$-$V$ by fluctuations and noise.}
        \label{IVRAC}
    \end{center}
\end{figure*}

In experiment it often happens that the $I$-$V$ is asymmetric with different positive and negative critical currents, $I_{c+} \neq I_{c-}$. This is typically due to the self-field effect, or junction inhomogeneity \cite{Krasnov_1997,Krasnov_2020}. In this case $(I_c/I_{ac})^k$ ($k=2,4$) in Eqs. (3) and (5) should be replaced by the mean value $[(I_{c+}/I_{ac})^k+(I_{c-}/I_{ac})^k]/2$. Since now there are two unknown parameters $I_{c+}$ and $I_{c-}$, their determination requires the knowledge of both $R_1$ and $R_3$: 
\begin{equation}
    \begin{split}
    I_{c\pm} = \frac{I_{ac}}{\sqrt{2}}\sqrt{a \pm \sqrt{b - a^2}}, \\
    a = \bigg(1-\frac{R_1}{R_n} \bigg),~ b = \bigg(\frac{R_3}{R_n} + \frac{R_1}{R_n} -1\bigg).
    \end{split}
\end{equation}
All even harmonics remain zero, unless there is a hysteresis in the $I$-$V$ with retrapping current $I_r<I_c$ \cite{Krasnov_2007}. 
{ In this case Eqs. (3) and (4) should be replaced by $R_1/R_n=(a_c^2+a_r^2)/2$ and $R_2/R_n=(4/3\pi)[a_c^3-a_r^3]$, where $a_{c,r}^2=1-(I_{c,r}/I_{ac})^2$. Similar to the asymmetric case, Eq.(8), measurements of two harmonics $R_{1,2}$ are needed for determination of the two unknown variables $I_c$ and $I_r$ in this case.}

Finally, we note that the shape of the $I$-$V$ may deviate from the RSJ expression, Eq. (1). In general, a similar analysis can be expanded to any shape of the $I$-$V$. We will not consider this rigorously because { there is no explicit} analytical solution. Instead, we propose a simple phenomenological modification of Eq. (6) with an additional fitting parameter $\beta$:
\begin{equation}\label{IcRac}
I_c(R_1)=I_{ac}\left[1-\frac{R_{1}}{R_n}\right]^{\beta},
\end{equation}
with $\beta=0.5$ in the RSJ case, Eq. (6).

\begin{figure*}[!ht]
    \begin{center}
        \includegraphics[width=14cm]{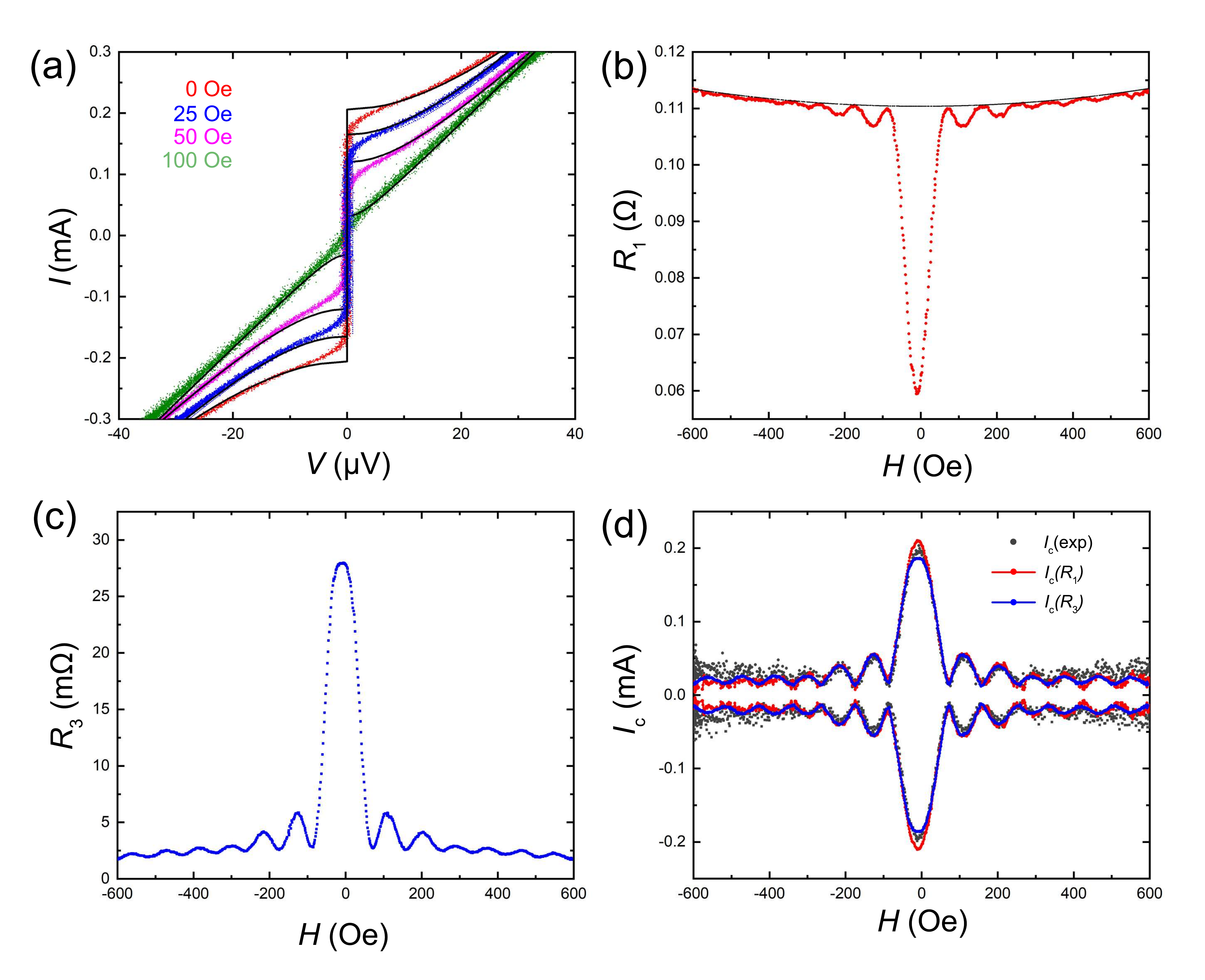}
        \caption{(a) $I$-$V$ characteristics of the Nb-PtNi-Nb junction at $T=4.47$ K and at different in-plane magnetic fields. Black lines represent RSJ fits. (b) and (c) Field modulation of the 1st (b) and the 3rd (c) lock-in harmonics of resistance for this junction. (d) Magnetic field modulation of { critical currents measured} experimentally, $I_c\text{(exp)}$, (black symbols) and reconstructed from the 1st (red) and 3rd (blue line) lock-in harmonics. }
        \label{fig2}
    \end{center}
\end{figure*}

\section{Comparison with experiment}
We present data for nano-scale proximity-coupled junctions Nb-PtNi-Nb and Nb CuNi-Nb. Junctions are made from trilayer films using 3D nanosculpturing by Focused Ion Beam (FIB). Details of fabrication and junction characteristics can be found in Refs. \cite{Iovan_2014, Kapran_2020, Kapran_2021}. Figure~\ref{IVRAC}(a) shows scanning electron microscopy (SEM) image and a sketch of one of the studied Nb-PtNi-Nb junctions (see Ref. \cite{Kapran_2021} for more details about properties of Nb-PtNi-Nb JJ's).  

Fig.~\ref{IVRAC}(b) shows the $I$-$V$ characteristics of a Nb-PtNi-Nb junction with sizes $250 \times 1000$ nm$^2$ at a fixed $T=4.47$ K and with no applied magnetic field, $H=0$. Red dots represent experimental data and a thin black line - corresponding numerical fits using the RSJ Eq. (1). It is seen that the fit is good with exception of the region close to $I_c$. The deviation may be either due to an intrinsic difference of the $I$-$V$ shape with a smooth{ er} than Eq. (1) increase of voltage at $I\simeq I_c$, or due to smearing by fluctuations and noise \cite{Kautz_1990,Krasnov_2007}. Therefore, the fit by Eq. (1) yields a somewhat overestimated value of $I_c\textrm{(Eq.1)}=200~\mu$A, which is larger than the value deduced from the experimental $I$-$V$, $I_c\textrm{(exp)}= 187 \pm 8~\mu$A, where the uncertainty is due to smearing.  

Fig.~\ref{IVRAC}(c) represents the measured 1st harmonic resistance, $R_1$, for this junction as a function of $I_{ac}$ (red circles) at $H=0$ and $T=4.47$ K. Lock-in measurements are performed at $f=13$ Hz { with the averaging time of 1 s}. The black solid line is obtained from Eq. (3), using $I_c$ as the only fitting parameter. The fit works well in a broad range of $I_{ac}$ and yields $I_c\textrm{(Eq.3)}=200~\mu$A. Fig. ~\ref{IVRAC}(d) represent $I_c$ deduced from the same $R_1(I_{ac})$ data with the help of Eq. (6), using $R_n$ as the only fitting parameter. Horizontal lines show $I_c\textrm{(exp)}$ (solid) and $I_c\textrm{(Eq.1)}=I_c\textrm{(Eq.3)}$ (dashed line) values. It can be seen that all methods of reconstruction of $I_c$ from $R_1$ work well and provide $I_c$ values in the range of experimental uncertainties, marked by error bars on $I_c\textrm{(exp)}$ in Fig.~\ref{IVRAC}(d). From Figs.~\ref{IVRAC}(c) and (d) it is seen that such the reconstruction provides reliable $I_c$ values in a broad bias range $1.3I_c <I_{ac}< 2I_c$. Discrepancies outside this range are caused by deviations of the $I$-$V$ shape from the RSJ Eq. (1) due to smearing at low bias and, possibly, self-heating at large bias \cite{Krasnov_2007}. The independence of the extracted $I_c$ from the bias, $I_{ac}$, indicates the robustness of the method.   

\begin{figure*}[ht!]
    \begin{center}
        \includegraphics[width=17cm]{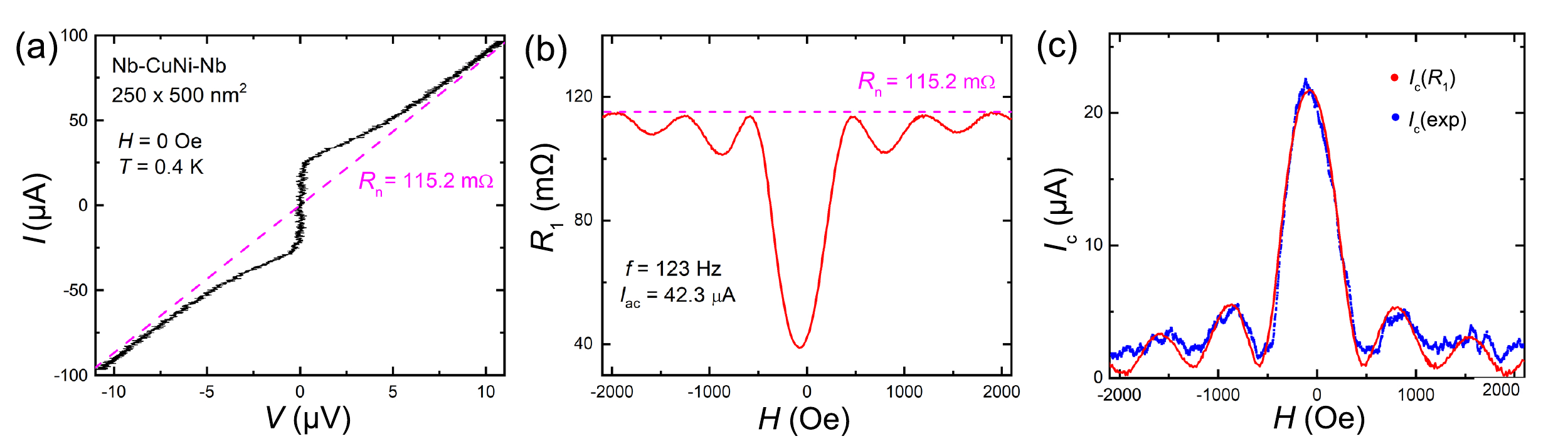}
        \caption{(a) The $I$-$V$ characteristics of a Nb-CuNi-Nb junction at $H=0$ and $T \simeq 0.4$ K. The dashed line indicates normal resistance $R_n=115.2$m$\Omega$. (b) Measured field dependence of the 1st harmonic lock-in resistance $R_1$. The horizontal dashed line indicates the $R_n$ level. (c) Field modulation of the critical current, determined from $I$-$V$'s (blue symbols), and recalculated from $R_{1}$ (red line) using Eq. (9).}
        \label{fig3}
    \end{center}
\end{figure*}

\section{Reconstruction of magnetic field modulation $I_c(H)$}

Magnetic field modulation, $I_c(H)$, is a figure of merit for JJ quality and uniformity \cite{Krasnov_1997}. Measurement of $I_c(H)$ at integer number of flux quanta in the JJ and at high fields, when $I_c(H)$ becomes small, is challenging because of enhanced susceptibility to fluctuations and noise at low Josephson energies \cite{Martinis_1987}. Lock-in measurements of $I_c$ become particularly useful in this case \cite{Iovan_2014,Kapran_2020}.   

Fig.~\ref{fig2}(a) shows a set of $I$-$V$'s for the same Nb-PtNi-Nb JJ at $T=4.47$ K and for different in-plane magnetic fields perpendicular to the long side of the JJ. It is seen that the $I_c$ is completely suppressed at $H\simeq 100$ Oe. Figs.~\ref{fig2}(b) and (c) show 1st and 3rd harmonics of lock-in resistance vs. $H$, measured at a fixed $I_{ac}=315~\mu$A. It is seen that both are carrying information about the Fraunhofer $I_c(H)$ modulation. 
Due to the small sizes of the JJ, the flux quantization field and the overall field range is rather large. This leads to a visible parabolic fields dependence of the junction resistance $R_n(H)$, indicated by the black line in Fig.~\ref{fig2}(b). Black dots in Fig.~\ref{fig2}(d) represent magnetic field modulation of $I_c\textrm{(exp)}$, obtained directly from the $I$-$V$'s. The determination is made using a threshold voltage criterion, $V<V_{th}$. Red and blue lines represent $I_c(R_1)$ and $I_c(R_3)$ values, recalculated from the 1st and 3rd lock-in harmonics, respectively, using Eqs. (6) and (7) with the actual $R_n(H)$ dependence, shown in Fig.~\ref{fig2}(b). It is seen that both modulation patterns $I_c(R_1)$ and $I_c(R_3)$ are in a quantitative agreement with  $I_c\textrm{(exp)}$ in the whole field range.  At high fields, $|H|>300$ Oe, modulation of $I_c\textrm{(exp)}$ is practically unresolvable, but for $I_c(R_1)$ and $I_c(R_3)$ it is clearly seen. Furthermore, $I_c(R_3)$ has a significantly larger signal-to-noise ratio than $I_c(R_1)$ due to a smaller $1/f$ noise.  

In Figure \ref{fig3} we analyze data for another Nb-CuNi-Nb junction with sizes $250 \times 500$ nm$^2$ (for more details about junction properties, see Refs. \cite{Iovan_2014,Kapran_2020}). Fig. \ref{fig3} (a) shows the $I$-$V$ at $H=0$ and $T \simeq 0.4$ K. Here a deviation from the RSJ shape, Eq. (1), in a form of a smoother, almost linear, deviation of $V$ from zero at $I\sim I_c$ can be seen more clearer than for the Nb-PtNi-Nb JJ, Fig. \ref{IVRAC} (a). Fig. \ref{fig3} (b) shows field modulation (for the downward field sweep) of the 1st harmonic lock-in resistance measured at $f=123$ Hz and $I_{ac}\simeq 42.3~\mu$A. Fig. \ref{fig3} (c) shows magnetic field modulation of the measured $I_c\textrm{(exp)}$ (blue symbols) obtained using a threshold criterion from the $I$-$V$ curves. Since the shape of $I$-$V$'s of this junction deviates from RSJ, we used the modified expression Eq. (9) using $\beta$ as the only fitting parameter for extraction of $I_c(R_1)$. The red line in Fig. \ref{fig3} (c) demonstrates a result of
such fitting with $\beta=0.8$. Apparently, it not only properly reproduces
$I_c(H)$, but also significantly reduces noise and corrects an
artifact of inaccurate dc-measurement of small critical currents,
$I_c < V_{th}/R_n$. Thus, introduction of a phenomenological parameter $\beta$ provides a simple way of accounting for the non-RSJ shape of the $I$-$V$ curve of a junction. 

{ Finally, we want to emphasize, that the discussed method is applicable for junctions with RSJ-like $I$-$V$'s with arbitrary $I_c$ and $R_n$, at any $T$, and for any type of fluctuations (quantum or thermal). In Fig. \ref{fig3} (c) the smallest reconstructed $I_c$ at $H\sim \pm 2000$ Oe is in the 100 nA range and the readout voltage $I_c R_n \sim 10$ nV. These are very good number for conventional measurements with averaging time of 1 s and without any special precautions. }

\section*{CONCLUSIONS}
To summarize, we have shown that lock-in measurements can be advantageously used for accurate determination of critical currents in small Josephson junctions, for which direct dc-determination of $I_c$ is complicated by noise and fluctuations. We have derived explicit and simple analytic expressions for the RSJ model and suggested a simple phenomenological modification for the non-RSJ case. The formalism was verified experimentally on nano-scale proximity-coupled junctions. We conclude that it is advantageous to measure both the 1st and the 3rd lock-in harmonics, which together allow robust and almost bias-{ in}dependent reconstruction of the critical current. Generally it may be useful also measure higher odd harmonics for further improvement of the proposed method. We argue that the developed technique provides a major advantage for read-out of various superconducting devices. 

\section*{Acknowledgments}
The work was supported by the Russian Science
Foundation grant No. 19-19-00594. We are grateful to A. Iovan for assistance with fabrication of Nb-CuNi-Nb junction. The manuscript was written during a sabbatical semester of V.M.K.
at MIPT, supported by the Faculty of Science at SU. 



\end{document}